\documentclass[a4paper,11pt]{article}  
\usepackage{aaskaiid}
\usepackage{orcidlink}

\title{SKA VLBI survey of the Southern sky for astrometry, geodesy, and astrophysics}
\ShortTitle{SKA VLBI survey}

\author[1]{M.~H.~Xu\orcidlink{0000-0001-9602-9489}}
\ShortName{Xu, Petrov, and Kovalev} 
\author[2]{L.~Y.~Petrov\orcidlink{0000-0001-9737-9667}}
\author[3]{Y.~Y.~Kovalev\orcidlink{0000-0001-9303-3263}}

\affiliation[1]{GFZ Helmholtz Centre for Geosciences, Telegrafenberg, 14476 Potsdam, Germany}
\emailAdd{minghui.xu@gfz.de}
\affiliation[2]{NASA Goddard Space Flight Center, Code 61A, 8800 Greenbelt Rd, Greenbelt, 20771 MD, USA}
\emailAdd{Leonid.Petrov@lpetrov.net}
\affiliation[3]{Max Planck Institute for Radio Astronomy, Auf dem Huegel 69, 53121 Bonn, Germany}
\emailAdd{yykovalev@gmail.com}

\abstract{
The development of SKA will open the opportunity to run dedicated VLBI
surveys of the Southern sky and observe sources not visible at
radio telescopes located in the Northern hemisphere. These surveys will
allow for doing geodesy with a Southern hemisphere-centered network, 
increase the density of compact radio sources that can be used as
calibrators, further extend the celestial reference frame  to deep south, and facilitate
high-precision differential astrometry for a wide range of applications.
Achieving a deep completeness level and determining the parsec-scale
properties of extragalactic radio sources is crucial for multi-wavelength
and multi-messenger astrophysics. This includes supporting Cherenkov
and neutrino telescope science cases, as well as joint VLBI-Gaia studies
of active galaxies. 
}

\begin{document}
\maketitle

\section{Introduction}

The most complete catalogue of compact radio sources derived
from analysis of VLBI observations at centimeter wavelengths, that is available so far, is 
the Radio Fundamental Catalogue \citep[RFC;][]{r:rfc1}, updated
on a quarterly basis and available at   
\href{https://doi.org/10.3847/1538-4365/ad8c36}{https://doi.org/10.3847/1538-4365/ad8c36}.
By October 2025 the catalogue contained 21,949 objects. The RFC is based on analysis of all publicly available observations from absolute astrometry and geodesy programs. However, the catalogue
has a significant disparity between declination zones $[-90^\circ, 
-40^\circ]$ and $[-40^\circ, +90^\circ]$: the density of compact radio sources listed in the RFC in the first zone is 644 objects per steradian versus 1988 objects in the second zone. Contribution to the RFC in the latter zone is almost exclusively based on observations from the Very Long Baseline Array (VLBA) located at the Northern hemisphere. The list of known sources with parsec scale-emission in $[-90^\circ, -40^\circ]$ originates from the VLBI geodetic program that started in 1980s (39 sources or 3\%),
CRF program \citep{r:fey04} (22 sources or 2\%), CRDS program \citep{r:crds}
(182 sources or 13\%), the Search for SOuthern Fermi Unassociated sources program
(SOFUS) (L.~Petrov, in preparation; 126 sources or 9\%), and the Long Baseline
Calibrator Surveys \citep{r:lcs1,r:lcs2} (1100 sources or 76\%). Among 1447 objects in the RFC at declinations $<-40^\circ$, 643 were detected at 2.3 and 8.4~GHz, and 804 objects (56\%) detected at 8.4~GHz only. The SKA VLBI survey based on deep radio surveys of the Southern sky \citep[e.g.,][]{2025PASA...42...38D,2011PASA...28..215N} will improve the completeness of radio calibrator catalogs such as the RFC.

In this chapter we discuss geodetic, astrometric, and astrophysical science cases as well as requirements for a deep targeted survey of compact extragalactic radio sources in the southern sky by an SKA VLBI system.

\section{Geodesy}

The technique of VLBI begun to be used for geodesy and astrometry in 1970, after its development in 1960s by astronomers for imaging radio sources. The first Southern hemisphere VLBI telescope was built at Hartebeesthoek, South Africa dedicated for geodesy in 1986, followed by the Hobart 26-meter telescope in Australia in 1989 and telescopes at O'Higgins, Antarctica and Fortaleza, Brazil in 1993. In 2010, three telescopes at Hobart, Katherine, and Yarragadee in Australia and one at Warkworth in New Zealand expanded the geodetic/astrometric VLBI network in the South. 

Since 2019, the new generation geodetic VLBI system \citep[see, e.g.,][]{VGOS_Niell} has started to operate regularly to make geodetic and astrometric observations, currently with a network of 16 to 19 telescopes observing 24 hours once per week. Due to the asymmetric distribution of the telescopes between North and South --- only four telescopes available in the South, the Southern telescopes have been less efficient than those in the North, e.g., the average number of observations per telescope is 1.5 to 4 times less for the southern telescopes. Given this asymmetric network, the geodetic/astrometric observations need to observe the radio sources in the Northern hemisphere more often than the South to acquire as many observables as possible. 

\begin{figure}
    \centering
    \captionsetup{justification=justified}
    \captionsetup{width=.95\linewidth}
    \includegraphics[width=0.85\linewidth]{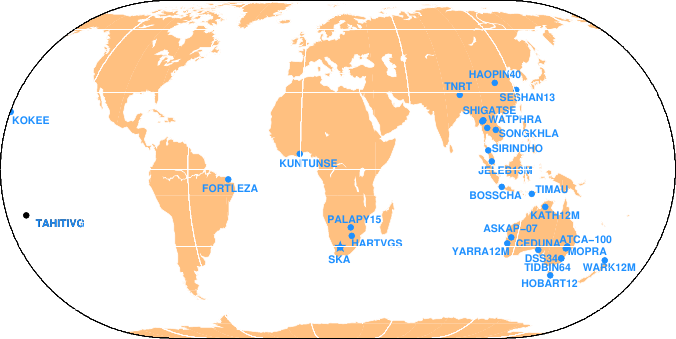}
    \caption{A Southern hemisphere centered network with SKA. At Tahiti site, indicated by a black dot, there is a plan to build a quad-band telescope as one of the geodetic stations while at the other sites radio telescopes are already available (either operating or under signal chain tests).}
    \label{fig:Southern_network}
\end{figure}

SKA will enable for doing geodesy/astrometry with a Southern hemisphere-centered observing program and an improved sensitivity to observe much weaker sources than the current geodetic network can do. Figure \ref{fig:Southern_network} shows the radio telescopes that may join the SKA for VLBI observations. It will allow for the following research and applications in geodesy:

\begin{enumerate}
        \item Determining the positions of the SKA telescopes in the global TRF and monitoring their velocities. This will allow for high accuracy differential astrometry with the SKA.
        \item Reducing the dominant impact of Northern stations on the scale of terrestrial reference frame (TRF). \citet{Kern:2025} claimed that the non-linear motion of VLBI station NYALES20 in Norway is responsible for more than 50\% of the drift in the TRF scale as reported in \citet{Altamimi:2023}. The impact of one specific VLBI station in the far North on the global TRF can be mitigated by a geodetic observing network in the South. 
        \item Measuring the relative positions of the telescopes of the SKA-mid AA configuration with sub-mm accuracy. It was demonstrated \citep[see, e.g.,][]{Xu:2023} that resolving phase delays on the short baselines allows for sub-mm accuracy of station positions. This can be done for 254 SKA stations to link the array together.
        \item Establishing the SKA array as the core geodetic station with respect to which the positions and motions of the other Southern telescopes are determined with higher accuracy. The TRF is established through geodetic observations dominated by the Northern hemisphere-centered network; there is no comparable observing network in the South so far. A regular observing program with the SKA as shown in Fig. \ref{fig:Southern_network} will improve the global geodesy.
        \item Investigating the potential usage of the radiometers for the tropospheric turbulence. The atmospheric turbulence on the time scales less than 1 hour (to a few seconds) introduces delay errors of tens of pico-seconds, and it causes one of the major challenges to improve the geodetic accuracy. The water vapor radiometers at the SKA site in real-time tracking of VLBI observations will provide valuable data to investigate how to model these errors.
\end{enumerate}

\section{Calibration of SKA with co-located GNSS receivers}\label{sec:GNSS}

An array of a size of 180~km is a factor of 50 times smaller than 
a global array with baseline projection lengths up to 9,000~km, 
which is close to the practical limit of effectively scheduling VLBI observations.
Considering the floor of the accuracy of VLBI absolute astrometry of 
0.1--0.2~mas due to systematic errors \citep{r:rfc1}, we can expect an accuracy 
of absolute astrometry of 5--10 mas or better from the SKA if we will take the 
systematic errors into account. 
One of the major factors that affect accuracy of geodesy and absolute 
astrometry observations is residual errors of modeling path delay
in the neutral atmosphere \citep[see, e.g.,][]{r:conc} after 
solving for zenith path delay from the same radio interferometry 
observations. Several recent studies demonstrated that the effect due to the angular structure 
of the radio sources is one of the major systematic error sources \citep[see, e.g.,][]{Anderson:2018,Xu:2021}. 

Source structure may have minimum effect on the observations from the SKA only. However, 
observations at an array of a size of 180~km are not favorable
for a reliable estimation of the tropospheric path delay: the differences
in elevations at different antennas do not exceed $1.5^\circ$, which 
makes estimates of zenith path delay and clock highly correlated. This 
high correlation is further extended to the parameter of station position 
in the astronomical observations, where the radio sources in a small
or regional sky area are observed. 
This problem can be mitigated by the emerging technique of micro-VLBI 
successfully demonstrated in \citet{r:microVLBI}. If a permanent Global 
Navigation Satellite System (GNSS) antenna is deployed at a distance 
of 70--100~m of a radio telescope that is capable to record voltage 
in a range of 1.17 to 1.66~GHz --- the frequency range of GPS, GLONASS, 
Galileo, and Beidou geodetic satellites, such an antenna can be used 
as an element of a radio interferometer with a radio telescope. 
Processing observations of GNSS satellites on an array consisting of GNSS 
antenna and radio telescopes allows us to determine fringe phase. After 
resolving phase delay ambiguities, the baseline vector connecting 
a radio telescope and a co-located GNSS antenna can be evaluated with 
a one millimeter accuracy. The feasibility of this approach has been 
proven by recent observations at VLBA and co-located
GNSS antennas deployed in 2024--2025 (J.~Skeens, in preparation).
Such observations will tie a radio telescope of the SKA array to 
the terrestrial coordinate system realized by the ground network 
of GNSS antennas with a millimeter level of accuracy. Implementation of 
such ties at {\it every SKA station} effectively 
transforms SKA to an enhanced SKA+GNSS array. Furthermore, analysis of GNSS data from co-located antennas 
allows us to derive zenith path delays regardless 
of SKA data. The capability to estimate zenith path delay and position of 
elements of SKA antennas from observations of GNSS satellites
using co-located GNSS antennas makes {\it a transformative impact} 
on the ability to derive absolute positions of radio sources. The method
of absolute astrometry implies a simultaneous determination of the
position and orientation of an instrument --- in this case the SKA+GNSS
array --- and positions of radio sources as well as atmospheric path delays.

This micro-VLBI technique has a potential to contribute to geodesy directly. Installation of a permanent
GNSS antenna on a deep drilled braced monument \citep[see, e.g.,]
[and reference therein]{r:ddbm} makes a geodetic network of companion GNSS antennas suitable for monitoring
changes in their positions due to tectonic motion and environmental
effects with a sub-millimeter level of accuracy. Regular observations of 
GNSS satellites using the micro-VLBI technique will provide independent 
datasets, in addition to the phase delays of the SKA observations (see previous section), 
to measure relative motions of SKA antennas at a one millimeter level 
of accuracy with respect to the reference frame based on positions of GNSS antennas. The
differences in the results from VLBI phase delays and the micro-VLBI 
technique can be investigated at the 1-mm accuracy level, which may provide understanding of the causes of 
the station position differences between group delay and phase delay.

Deployment of a permanent GNSS station within 70--100 meters away from each SKA
radio telescope provides another advantage. Observations of dual- 
and triple- band GNSS satellites allows us to determine slant 
ionospheric path delays for calibrating the SKA observations with a single frequency. Nowadays, typically 
40 to 50 satellites are visible at a given site and a given moment of 
time. Processing slant path delays due to ionosphere to 40 to 50 satellites from 254 
planned SKA stations provides a large dataset that can be used 
for determination of time-varying biases in the Global Ionospheric 
Model (GIM). It has been demonstrated by \citet{r:sba} that applying 
time-variable biases improves determination of the ionospheric path 
delay up to a factor of 3 with respect of the use of the GIM derived from 
analysis of a global network of GNSS antennas. This approach of 
refining GIM using slant ionospheric delay from co-located GNSS 
antennas can be further extended to a development of a regional 
ionospheric model that will help to mitigate the impact 
of residual ionospheric delays on SKA astrometry in the case of 
astronomical observations with a single frequency.

\section{Astrometry}

In the context of this article, astrometry refers to absolute astrometry that uses
the observable of the total delay derived from the cross correlation of the radio signals
to determine positions of radio sources, station positions, Earth orientation
parameters, and some calibration parameters simultaneously in a single solution.
In contrast, differential astrometry uses the observable of the differential (phase) delay
among the objects (including targets and calibrators) in a small sky area to mitigate the common systematic errors and determines their position differences --- the positions of the targets can thus be derived if the positions of the calibrators have been determined from absolute astrometry programs. 

For the high accuracy astrometry beyond the current capability, it requires modeling the effects due to structure of the radio sources in group delay observables because source structure causes astrometric positions to change along the jet direction over time and over frequency \citep[e.g.,][]{r:rfc1,Xu:2025}. This eventually brings geodesy, astrometry, and astrophysics together. For the same VLBI experiment, radio images can be derived and then used as source models in the geodetic data processing to derive astrometric positions that are free of these effects. It will allow for directly registering the absolute astrometric position on the radio image, and thus for the identification of the location of the optical emission as measured by Gaia. In these VLBI observations, a limited (but reasonable) number of sources can be observed in each 24-hour experiment to have a reasonably good sky coverage every one hour for estimating tropospheric delays and to have a good uv coverage for each source to obtain the images. 

\section{Astrophysics}
\label{s:astrophysics}

There are numerous astrophysical applications that require deep, complete, all-sky samples of VLBI-selected extragalactic radio sources, or a high density of compact calibrators enabling phase-referencing observations of weak targets. These include:
\begin{enumerate}
\item Reconstruction of the Galactic scattering screen properties \citep{2022MNRAS.515.1736K};
\item Studies of VLBI–Gaia position offsets, which trace synchrotron opacity effects and electron re-acceleration sites in relativistic jets \citep{2017A&A...598L...1K,2019ApJ...871..143P,2025A&A...693A.270F};
\item Identification and investigation of Compact Symmetric Objects \citep[e.g.,][]{2024ApJ...961..240K};
\item Identification and characterization of AGN producing high and very-high-energy $\gamma$-rays \citep[e.g.,][]{2009ApJ...707L..56K,r:rfc1,2018ApJ...853...68P};
\item Identification and study of potential astrophysical neutrino sources \citep[e.g.,][]{2023MNRAS.523.1799P,2025A&A...700L..12K};
\item A search for and characterization of new strong gravitational lenses \citep{2010ARA&A..48...87T,2015aska.confE..84M};
\item Precision differential astrometry, including pulsar and maser parallax measurements, time-domain VLBI, and transient sky studies requiring phase-reference calibrators \citep[e.g.,][]{2014ARA&A..52..339R,2019ApJ...875..100D,2020Natur.577..190M}.
\end{enumerate}

\section{Summary}

VLBI observations with the SKA, supported by the co-located GNSS antennas and water vapor radiometers, allow for a wide variety of scientific applications for geodesy, astrometry, and astrophysics. The SKA will open unique possibilities in these fields because of its geographical location and the high sensitivity. 

Global geodetic and astrometric VLBI observations with the SKA need to be carried to tie the positions of the SKA telescopes to the global terrestrial reference frame and measure the positions of the radio sources for the improved accuracy. To measure ionospheric-free group delay precisely that is needed for geodesy and astrometry, two or more frequency bands with separation of a factor of larger than 1.7 and with a broad bandwidth per band ($>$ 0.5 GHz) should be used.
%


\bigskip
%
MHX was supported by the Astrogeodesy project (ERC grant agreement No 101076060) and YYK was supported by the MuSES project (ERC grant agreement No 101142396), both of which have received funding from the European Union. Views and opinions expressed are however those of the author(s) only and do not necessarily reflect those of the European Union or ERCEA. Neither the European Union nor the granting authority can be held responsible for them.
This research has made use of the NASA/IPAC Extragalactic Database, which is funded by the National Aeronautics and Space Administration and operated by the California Institute of Technology.


\bibliographystyle{abbrvnat-maxbibnames4}
\newcommand{\actaa}{Acta Astron.} 
\newcommand{\araa}{ARA\&A} 
\newcommand{\aar}{A\&ARv} 
\newcommand{\aapr}{A\&ARv} 
\newcommand{\ab}{Astrobiol.} 
\newcommand{\aj}{AJ} 
\newcommand{\apj}{ApJ} 
\newcommand{\apjl}{ApJL} 
\newcommand{\apjs}{ApJSS} 
\newcommand{\ao}{Appl. Opt.} 
\newcommand{\apss}{Astro. \& Space Sci.} 
\newcommand{\aap}{A\&A} 
\newcommand{\aaps}{A\&AS.} 
\newcommand{\baas}{Bull. Am. Astron. Soc.} 
\newcommand{\caa}{Chinese A\&A} 
\newcommand{\cjaa}{Chinese J. A\&A} 
\newcommand{\cqg}{Class. Quantum Gravity} 
\newcommand{\gal}{Galaxies} 
\newcommand{\gca}{Geo. Cosmo. Acta} 
\newcommand{\icarus}{Icarus} 
\newcommand{\jcap}{JCAP} 
\newcommand{\jgr}{J. Geophys. Res.} 
\newcommand{\jgrp}{J. Geophys. Res. Planets} 
\newcommand{\jqsrt}{J. Quant. Spectrosc. Radiat. Transf.} 
\newcommand{\memsai}{Mem. SAIt} 
\newcommand{\mnras}{MNRAS} 
\newcommand{\nat}{Nature} 
\newcommand{\nastro}{Nat. Astron.} 
\newcommand{\ncomms}{Nat. Commun.} 
\newcommand{\nphys}{Nat. Phys.} 
\newcommand{\na}{New Astron.} 
\newcommand{\nar}{New Astron. Rev.} 
\newcommand{\physrep}{Phys. Rep.} 
\newcommand{\pra}{Phys. Rev. A} 
\newcommand{\prb}{Phys. Rev. B} 
\newcommand{\prc}{Phys. Rev. C} 
\newcommand{\prd}{Phys. Rev. D} 
\newcommand{\pre}{Phys. Rev. E} 
\newcommand{\prx}{Phys. Rev. X} 
\newcommand{\prl}{Phys. Rev. Let.} 
\newcommand{\psj}{Planet. Sci. J.} 
\newcommand{\planss}{Planet. Space Sci.} 
\newcommand{\pnas}{Proc. Natl Acad. Sci. USA} 
\newcommand{\procspie}{Proc. SPIE} 
\newcommand{\pasa}{PASA} 
\newcommand{\pasj}{PASJ} 
\newcommand{\pasp}{PASP} 
\newcommand{\rmxaa}{RMXAA} 
\newcommand{\sci}{Science} 
\newcommand{\sciadv}{Sci. Adv.} 
\newcommand{\solphys}{Sol. Phys.} 
\newcommand{\sovast}{Soviet Ast.} 
\newcommand{\ssr}{Space Sci. Rev.} 
\newcommand{\uni}{Universe} 

\setlength{\bibsep}{0.0pt} 
\bibliography{vlbi_survey}

@ARTICLE{2010ARA&A..48...87T,
       author = {{Treu}, Tommaso},
        title = "{Strong Lensing by Galaxies}",
      journal = {\araa},
         year = 2010,
        month = sep,
       volume = {48},
        pages = {87-125},
          doi = {10.1146/annurev-astro-081309-130924},
archivePrefix = {arXiv},
       eprint = {1003.5567},
 primaryClass = {astro-ph.CO},
       adsurl = {https://ui.adsabs.harvard.edu/abs/2010ARA&A..48...87T}
}

@INPROCEEDINGS{2015aska.confE..84M,
       author = {{McKean}, J. and {Jackson}, N. and {Vegetti}, S. and {Rybak}, M. and {Serjeant}, S. and {Koopmans}, L.~V.~E. and {Metcalf}, R.~B. and {Fassnacht}, C. and {Marshall}, P.~J. and {Pandey-Pommier}, M.},
        title = "{Strong Gravitational Lensing with the SKA}",
    booktitle = {Advancing Astrophysics with the Square Kilometre Array (AASKA14)},
         year = 2015,
        month = apr,
          eid = {84},
        pages = {84},
          doi = {10.22323/1.215.0084},
archivePrefix = {arXiv},
       eprint = {1502.03362},
 primaryClass = {astro-ph.GA},
       adsurl = {https://ui.adsabs.harvard.edu/abs/2015aska.confE..84M}
}

@ARTICLE{2020Natur.577..190M,
       author = {{Marcote}, B. and {Nimmo}, K. and {Hessels}, J.~W.~T. and {Tendulkar}, S.~P. and {Bassa}, C.~G. and {Paragi}, Z. and {Keimpema}, A. and {Bhardwaj}, M. and {Karuppusamy}, R. and {Kaspi}, V.~M. and {Law}, C.~J. and {Michilli}, D. and {Aggarwal}, K. and {Andersen}, B. and {Archibald}, A.~M. and {Bandura}, K. and {Bower}, G.~C. and {Boyle}, P.~J. and {Brar}, C. and {Burke-Spolaor}, S. and {Butler}, B.~J. and {Cassanelli}, T. and {Chawla}, P. and {Demorest}, P. and {Dobbs}, M. and {Fonseca}, E. and {Giri}, U. and {Good}, D.~C. and {Gourdji}, K. and {Josephy}, A. and {Kirichenko}, A. Yu. and {Kirsten}, F. and {Landecker}, T.~L. and {Lang}, D. and {Lazio}, T.~J.~W. and {Li}, D.~Z. and {Lin}, H.-H. and {Linford}, J.~D. and {Masui}, K. and {Mena-Parra}, J. and {Naidu}, A. and {Ng}, C. and {Patel}, C. and {Pen}, U.-L. and {Pleunis}, Z. and {Rafiei-Ravandi}, M. and {Rahman}, M. and {Renard}, A. and {Scholz}, P. and {Siegel}, S.~R. and {Smith}, K.~M. and {Stairs}, I.~H. and {Vanderlinde}, K. and {Zwaniga}, A.~V.},
        title = "{A repeating fast radio burst source localized to a nearby spiral galaxy}",
      journal = {\nat},
         year = 2020,
        month = jan,
       volume = {577},
       number = {7789},
        pages = {190-194},
          doi = {10.1038/s41586-019-1866-z},
archivePrefix = {arXiv},
       eprint = {2001.02222},
 primaryClass = {astro-ph.HE},
       adsurl = {https://ui.adsabs.harvard.edu/abs/2020Natur.577..190M}
}

@ARTICLE{2019ApJ...875..100D,
       author = {{Deller}, A.~T. and {Goss}, W.~M. and {Brisken}, W.~F. and {Chatterjee}, S. and {Cordes}, J.~M. and {Janssen}, G.~H. and {Kovalev}, Y.~Y. and {Lazio}, T.~J.~W. and {Petrov}, L. and {Stappers}, B.~W. and {Lyne}, A.},
        title = "{Microarcsecond VLBI Pulsar Astrometry with PSR{\ensuremath{\pi}} II. Parallax Distances for 57 Pulsars}",
      journal = {\apj},
         year = 2019,
        month = apr,
       volume = {875},
       number = {2},
          eid = {100},
        pages = {100},
          doi = {10.3847/1538-4357/ab11c7},
archivePrefix = {arXiv},
       eprint = {1808.09046},
 primaryClass = {astro-ph.IM},
       adsurl = {https://ui.adsabs.harvard.edu/abs/2019ApJ...875..100D}
}

@ARTICLE{2014ARA&A..52..339R,
       author = {{Reid}, M.~J. and {Honma}, M.},
        title = "{Microarcsecond Radio Astrometry}",
      journal = {\araa},
         year = 2014,
       volume = {52},
        pages = {339-372},
          doi = {10.1146/annurev-astro-081913-040006},
archivePrefix = {arXiv},
       eprint = {1312.2871},
 primaryClass = {astro-ph.IM},
       adsurl = {https://ui.adsabs.harvard.edu/abs/2014ARA&A..52..339R}
}

@ARTICLE{2024ApJ...961..240K,
       author = {{Kiehlmann}, S. and {Lister}, M.~L. and {Readhead}, A.~C.~S. and {Liodakis}, I. and {O'Neill}, Sandra and {Pearson}, T.~J. and {Sheldahl}, Evan and {Siemiginowska}, Aneta and {Tassis}, K. and {Taylor}, G.~B. and {Wilkinson}, P.~N.},
        title = "{Compact Symmetric Objects. I. Toward a Comprehensive Bona Fide Catalog}",
      journal = {\apj},
         year = 2024,
        month = feb,
       volume = {961},
       number = {2},
          eid = {240},
        pages = {240},
          doi = {10.3847/1538-4357/ad0c56},
archivePrefix = {arXiv},
       eprint = {2303.11357},
 primaryClass = {astro-ph.HE},
       adsurl = {https://ui.adsabs.harvard.edu/abs/2024ApJ...961..240K}
}

@ARTICLE{2019ApJ...871..143P,
       author = {{Plavin}, A.~V. and {Kovalev}, Y.~Y. and {Petrov}, L.~Y.},
        title = "{Dissecting the AGN Disk-Jet System with Joint VLBI-Gaia Analysis}",
      journal = {\apj},
         year = 2019,
          doi = {10.3847/1538-4357/aaf650},
archivePrefix = {arXiv},
       eprint = {1808.05115},
 primaryClass = {astro-ph.GA},
       adsurl = {https://ui.adsabs.harvard.edu/abs/2019ApJ...871..143P}
}

@ARTICLE{2022MNRAS.515.1736K,
       author = {{Koryukova}, T.~A. and {Pushkarev}, A.~B. and {Plavin}, A.~V. and {Kovalev}, Y.~Y.},
        title = "{Tracing Milky Way scattering by compact extragalactic radio sources}",
      journal = {\mnras},
         year = 2022,
        month = sep,
       volume = {515},
       number = {2},
        pages = {1736-1750},
          doi = {10.1093/mnras/stac1898},
archivePrefix = {arXiv},
       eprint = {2201.04359},
 primaryClass = {astro-ph.GA},
       adsurl = {https://ui.adsabs.harvard.edu/abs/2022MNRAS.515.1736K}
}

@ARTICLE{2018ApJ...853...68P,
       author = {{Piner}, B. Glenn and {Edwards}, Philip G.},
        title = "{Multi-epoch VLBA Imaging of 20 New TeV Blazars: Apparent Jet Speeds}",
      journal = {\apj},
         year = 2018,
        month = jan,
       volume = {853},
       number = {1},
          eid = {68},
        pages = {68},
          doi = {10.3847/1538-4357/aaa425},
archivePrefix = {arXiv},
       eprint = {1801.00817},
 primaryClass = {astro-ph.HE},
       adsurl = {https://ui.adsabs.harvard.edu/abs/2018ApJ...853...68P}
}

@ARTICLE{2009ApJ...707L..56K,
       author = {{Kovalev}, Y.~Y.},
        title = "{Identification of the Early Fermi/LAT Gamma-Ray Bright Objects with Extragalactic VLBI Sources}",
      journal = {\apjl},
         year = 2009,
        month = dec,
       volume = {707},
       number = {1},
        pages = {L56-L59},
          doi = {10.1088/0004-637X/707/1/L56},
archivePrefix = {arXiv},
       eprint = {0908.4152},
 primaryClass = {astro-ph.CO},
       adsurl = {https://ui.adsabs.harvard.edu/abs/2009ApJ...707L..56K}
}

@ARTICLE{2017A&A...598L...1K,
       author = {{Kovalev}, Y.~Y. and {Petrov}, L. and {Plavin}, A.~V.},
        title = "{VLBI-Gaia offsets favor parsec-scale jet direction in active galactic nuclei}",
      journal = {\aap},
         year = 2017,
       volume = {598},
          eid = {L1},
        pages = {L1},
          doi = {10.1051/0004-6361/201630031},
archivePrefix = {arXiv},
       eprint = {1611.02632},
 primaryClass = {astro-ph.GA},
       adsurl = {https://ui.adsabs.harvard.edu/abs/2017A&A...598L...1K}
}

@ARTICLE{2025A&A...700L..12K,
       author = {{Kovalev}, Y.~Y. and {Pushkarev}, A.~B. and {G{\'o}mez}, J.~L. and {Homan}, D.~C. and {Lister}, M.~L. and {Livingston}, J.~D. and {Pashchenko}, I.~N. and {Plavin}, A.~V. and {Savolainen}, T. and {Troitsky}, S.~V.},
        title = "{Looking into the jet cone of the neutrino-associated very high-energy blazar PKS 1424+240}",
      journal = {\aap},
         year = 2025,
        month = aug,
       volume = {700},
          eid = {L12},
        pages = {L12},
          doi = {10.1051/0004-6361/202555400},
archivePrefix = {arXiv},
       eprint = {2504.09287},
 primaryClass = {astro-ph.HE},
       adsurl = {https://ui.adsabs.harvard.edu/abs/2025A&A...700L..12K}
}

@ARTICLE{2023MNRAS.523.1799P,
       author = {{Plavin}, A.~V. and {Kovalev}, Y.~Y. and {Kovalev}, Yu A. and {Troitsky}, S.~V.},
        title = "{Growing evidence for high-energy neutrinos originating in radio blazars}",
      journal = {\mnras},
         year = 2023,
        month = aug,
       volume = {523},
       number = {2},
        pages = {1799-1808},
          doi = {10.1093/mnras/stad1467},
archivePrefix = {arXiv},
       eprint = {2211.09631},
 primaryClass = {astro-ph.HE},
       adsurl = {https://ui.adsabs.harvard.edu/abs/2023MNRAS.523.1799P}
}

@ARTICLE{2025PASA...42...38D,
       author = {{Duchesne}, S. and {Ross}, K. and {Thomson}, A.~J.~M. and {Lenc}, E. and {Murphy}, T. and {Galvin}, T.~J. and {Hotan}, A.~W. and {Moss}, V.~A. and {Whiting}, M.~T.},
        title = "{The Rapid ASKAP Continuum Survey (RACS) VI: The RACS-high 1655.5 MHz images and catalogue.}",
      journal = {\pasa},
         year = 2025,
        month = jan,
       volume = {42},
          eid = {38},
        pages = {38},
          doi = {10.1017/pasa.2025.2},
archivePrefix = {arXiv},
       eprint = {2501.04978},
 primaryClass = {astro-ph.GA},
       adsurl = {https://ui.adsabs.harvard.edu/abs/2025PASA...42...38D}
}

@ARTICLE{2011PASA...28..215N,
       author = {{Norris}, Ray P. and {Hopkins}, A.~M. and {Afonso}, J. and {Brown}, S. and {Condon}, J.~J. and {Dunne}, L. and {Feain}, I. and {Hollow}, R. and {Jarvis}, M. and {Johnston-Hollitt}, M. and {Lenc}, E. and {Middelberg}, E. and {Padovani}, P. and {Prandoni}, I. and {Rudnick}, L. and {Seymour}, N. and {Umana}, G. and {Andernach}, H. and {Alexander}, D.~M. and {Appleton}, P.~N. and {Bacon}, D. and {Banfield}, J. and {Becker}, W. and {Brown}, M.~J.~I. and {Ciliegi}, P. and {Jackson}, C. and {Eales}, S. and {Edge}, A.~C. and {Gaensler}, B.~M. and {Giovannini}, G. and {Hales}, C.~A. and {Hancock}, P. and {Huynh}, M.~T. and {Ibar}, E. and {Ivison}, R.~J. and {Kennicutt}, R. and {Kimball}, Amy E. and {Koekemoer}, A.~M. and {Koribalski}, B.~S. and {L{\'o}pez-S{\'a}nchez}, {\'A}. R. and {Mao}, M.~Y. and {Murphy}, T. and {Messias}, H. and {Pimbblet}, K.~A. and {Raccanelli}, A. and {Randall}, K.~E. and {Reiprich}, T.~H. and {Roseboom}, I.~G. and {R{\"o}ttgering}, H. and {Saikia}, D.~J. and {Sharp}, R.~G. and {Slee}, O.~B. and {Smail}, Ian and {Thompson}, M.~A. and {Urquhart}, J.~S. and {Wall}, J.~V. and {Zhao}, G.-B.},
        title = "{EMU: Evolutionary Map of the Universe}",
      journal = {\pasa},
         year = 2011,
        month = aug,
       volume = {28},
       number = {3},
        pages = {215-248},
          doi = {10.1071/AS11021},
archivePrefix = {arXiv},
       eprint = {1106.3219},
 primaryClass = {astro-ph.CO},
       adsurl = {https://ui.adsabs.harvard.edu/abs/2011PASA...28..215N}
}

@ARTICLE{r:microVLBI,
       author = {{Skeens}, J. and {York}, J. and {Petrov}, L. and {Munton}, D. and {Herrity}, K. and {Ji-Cathriner}, R. and {Bettadpur}, S. and {Gaussiran}, T.},
        title = "{First Observations With a GNSS Antenna to Radio Telescope Interferometer}",
      journal = {Radio Science},
         year = 2023,
        month = aug,
       volume = {58},
       number = {8},
          eid = {e2023RS007734},
        pages = {e2023RS007734},
          doi = {10.1029/2023RS007734},
archivePrefix = {arXiv},
       eprint = {2304.11016},
 primaryClass = {physics.geo-ph},
       adsurl = {https://ui.adsabs.harvard.edu/abs/2023RaSc...5807734S}
}

@ARTICLE{r:sba,
       author = {{Petrov}, Leonid},
        title = "{Single-band VLBI Absolute Astrometry}",
      journal = {\aj},
         year = 2023,
        month = apr,
       volume = {165},
       number = {4},
          eid = {183},
        pages = {183},
          doi = {10.3847/1538-3881/acc174},
archivePrefix = {arXiv},
       eprint = {2211.04647},
 primaryClass = {astro-ph.IM},
       adsurl = {https://ui.adsabs.harvard.edu/abs/2023AJ....165..183P}
}

@ARTICLE{r:conc,
       author = {{Petrov}, Leonid and {Ploetz}, Christian and {Schartner}, Matthias},
        title = "{Assessment of the Earth orientation parameter accuracy from concurrent VLBI observations}",
      journal = {arXiv e-prints},
         year = 2025,
        month = jun,
          eid = {arXiv:2506.15859},
        pages = {arXiv:2506.15859},
          doi = {10.48550/arXiv.2506.15859},
archivePrefix = {arXiv},
       eprint = {2506.15859},
 primaryClass = {physics.geo-ph},
       adsurl = {https://ui.adsabs.harvard.edu/abs/2025arXiv250615859P}
}

@article{r:ddbm,
  title = {Performance analysis of geodetic monuments on GNSS time series noise property and velocity uncertainty estimation},
  volume = {76},
  DOI = {10.1016/j.asr.2025.08.036},
  number = {10},
  journal = {ASR},
  publisher = {Elsevier BV},
  author = {He,  Xiaoxing and Lv,  Hongli and Wang,  Shengdao and Ren,  Xiaodong and Yang,  Ronghua and Hu,  Shunqiang and Wang,  Bin and Chen,  Yanying},
  year = {2025},
  month = nov,
  pages = {5872–5889}
}

@ARTICLE{r:rfc1,
       author = {{Petrov}, L.~Y. and {Kovalev}, Y.~Y.},
        title = "{The Radio Fundamental Catalog. I. Astrometry}",
      journal = {\apjs},
         year = 2025,
        month = feb,
       volume = {276},
       number = {2},
          eid = {38},
        pages = {51},
          doi = {10.3847/1538-4365/ad8c36},
       adsurl = {https://ui.adsabs.harvard.edu/abs/2025ApJS..276...38P}
}

@ARTICLE{r:fey04,
       author = {{Fey}, Alan L. and {Ojha}, Roopesh and {Jauncey}, David L. and {Johnston}, Kenneth J. and {Reynolds}, John E. and {Lovell}, James E.~J. and {Tzioumis}, Anastasios K. and {Quick}, Jonathan F.~H. and {Nicolson}, George D. and {Ellingsen}, Simon P. and {McCulloch}, Peter M. and {Koyama}, Yasuhiro},
        title = "{Accurate Astrometry of 22 Southern Hemisphere Radio Sources}",
      journal = {\aj},
         year = 2004,
        month = mar,
       volume = {127},
       number = {3},
        pages = {1791-1795},
          doi = {10.1086/381957},
       adsurl = {https://ui.adsabs.harvard.edu/abs/2004AJ....127.1791F}
}

@ARTICLE{r:lcs1,
   author = {{Petrov}, L. and {Phillips}, C. and {Bertarini}, A. and {Murphy}, T. and 
	{Sadler}, E.~M.},
    title = "{The LBA Calibrator Survey of southern compact extragalactic radio sources - LCS1}",
  journal = {\mnras},
archivePrefix = "arXiv",
   eprint = {1012.2607},
 primaryClass = "astro-ph.CO",
     year = 2011,
    month = jul,
   volume = 414,
    pages = {2528-2539},
      doi = {10.1111/j.1365-2966.2011.18570.x},
   adskey = {2011MNRAS.414.2528P},
   adsurl = {http://adsabs.harvard.edu/abs/2011MNRAS.414.2528P}
}

@ARTICLE{r:lcs2,
       author = {{Petrov}, Leonid and {de Witt}, Alet and {Sadler}, Elaine M. and
         {Phillips}, Chris and {Horiuchi}, Shinji},
        title = "{The Second LBA Calibrator Survey of southern compact extragalactic radio sources - LCS2}",
      journal = {\mnras},
         year = "2019",
        month = "May",
       volume = {485},
       number = {1},
        pages = {88-101},
          doi = {10.1093/mnras/stz242},
archivePrefix = {arXiv},
       eprint = {1812.02916},
 primaryClass = {astro-ph.IM},
       adsurl = {https://ui.adsabs.harvard.edu/abs/2019MNRAS.485...88P}
}

@ARTICLE{r:crds,
       author = {{Weston}, S. and {de Witt}, A. and {Kr{\'a}sn{\'a}}, Hana and {Le Bail}, Karine and {Hardin}, Sara and {Gordon}, David and {Fengchun}, Shu and {Fey}, Alan and {Schartner}, Matthias and {Basu}, Sayan and {Titov}, Oleg and {Behrend}, Dirk and {Jacobs}, Christopher S. and {Hankey}, Warren and {Salguero}, Federico and {Reynolds}, John E.},
        title = "{On more than two decades of Celestial Reference Frame VLBI observations in the deep south: IVS-CRDS (1995-2021)}",
      journal = {\pasa},
         year = 2023,
        month = sep,
       volume = {40},
          eid = {e041},
        pages = {e041},
          doi = {10.1017/pasa.2023.33},
archivePrefix = {arXiv},
       eprint = {2306.06830},
 primaryClass = {astro-ph.EP},
       adsurl = {https://ui.adsabs.harvard.edu/abs/2023PASA...40...41W}
}

@ARTICLE{Kern:2025,
       author = {{Kern}, Lisa and {Kr{\'a}sn{\'a}}, Hana and {Nothnagel}, Axel and {B{\"o}hm}, Johannes},
        title = "{Terrestrial reference frame scale drift anomalies in VLBI and the contribution of Ny-{\r{A}}lesund radio telescopes}",
      journal = {Earth, Planets and Space},
         year = 2025,
        month = mar,
       volume = {77},
       number = {1},
          eid = {40},
        pages = {40},
          doi = {10.1186/s40623-025-02159-z},
       adsurl = {https://ui.adsabs.harvard.edu/abs/2025EP&S...77...40K}
}

@ARTICLE{Altamimi:2023,
       author = {{Altamimi}, Zuheir and {Rebischung}, Paul and {Collilieux}, Xavier and {M{\'e}tivier}, Laurent and {Chanard}, Kristel},
        title = "{ITRF2020: an augmented reference frame refining the modeling of nonlinear station motions}",
      journal = {Journal of Geodesy},
         year = 2023,
        month = may,
       volume = {97},
       number = {5},
          eid = {47},
        pages = {47},
          doi = {10.1007/s00190-023-01738-w},
       adsurl = {https://ui.adsabs.harvard.edu/abs/2023JGeod..97...47A}
}

@ARTICLE{Xu:2023,
       author = {{Xu}, Ming H. and {Savolainen}, Tuomas and {Bolotin}, Sergei and {Bernhart}, Simone and {Pl{\"o}tz}, Christian and {Haas}, R{\"u}diger and {Varenius}, Eskil and {Wang}, Guangli and {McCallum}, Jamie and {Heinkelmann}, Robert and {Lunz}, Susanne and {Schuh}, Harald and {Zubko}, Nataliya and {Kareinen}, Niko},
        title = "{Baseline Vector Repeatability at the Sub-Millimeter Level Enabled by Radio Interferometer Phase Delays of Intra-Site Baselines}",
      journal = {JGR-SE},
         year = 2023,
       volume = {128},
       number = {3},
          doi = {10.1029/2022JB02519810.1002/essoar.10512041.1},
       adsurl = {https://ui.adsabs.harvard.edu/abs/2023JGRB..12825198X}
}

@article{Xu:2025,
doi = {10.3847/1538-3881/adb133},
year = {2025},
month = {feb},
publisher = {The American Astronomical Society},
volume = {169},
number = {3},
pages = {173},
author = {Xu, Ming Hui and Charlot, Patrick},
title = {Variations of Absolute Source Positions Determined from Quad-band VLBI Observations},
journal = {AJ},
}

@ARTICLE{Anderson:2018,
       author = {{Anderson}, James M. and {Xu}, Ming H.},
        title = "{Source Structure and Measurement Noise Are as Important as All Other Residual Sources in Geodetic VLBI Combined}",
      journal = {JGR-SE},
         year = 2018,
       volume = {123},
       number = {11},
        pages = {10,162-10,190},
          doi = {10.1029/2018JB015550},
       adsurl = {https://ui.adsabs.harvard.edu/abs/2018JGRB..12310162A}
}

@ARTICLE{Xu:2021,
       author = {{Xu}, Ming H. and {Anderson}, James M. and {Heinkelmann}, Robert and {Lunz}, Susanne and {Schuh}, Harald and {Wang}, Guangli},
        title = "{Observable quality assessment of broadband very long baseline interferometry system}",
      journal = {Journal of Geodesy},
         year = 2021,
        month = may,
       volume = {95},
       number = {5},
          eid = {51},
        pages = {51},
          doi = {10.1007/s00190-021-01496-7},
archivePrefix = {arXiv},
       eprint = {2102.12750},
 primaryClass = {astro-ph.IM},
       adsurl = {https://ui.adsabs.harvard.edu/abs/2021JGeod..95...51X}
}

@ARTICLE{2025A&A...693A.270F,
       author = {{Fichet de Clairfontaine}, G. and {Perucho}, M. and {Mart{\'\i}}, J.~M. and {Kovalev}, Y.~Y.}, 
        title = "{Dynamic and radiative implications of jet{\textendash}star interactions in AGN jets}",
      journal = {\aap},
         year = 2025,
        month = jan,
       volume = {693},
          eid = {A270},
        pages = {A270},
          doi = {10.1051/0004-6361/202451914},
archivePrefix = {arXiv},
       eprint = {2412.07945},
 primaryClass = {astro-ph.HE},
       adsurl = {https://ui.adsabs.harvard.edu/abs/2025A&A...693A.270F}
}
\end{document}